\documentclass[twocolumn,aps,showpacs,superscriptaddress,floatfix]
{revtex4}
\usepackage[dvips]{graphicx}
\newcommand{\ket}[1]{\left | #1 \right \rangle}
\newcommand{\bra}[1]{\left \langle #1 \right |}

\newcommand{\tr}{{\rm \, Tr }\, }
\newcommand{\bm}[1]{\mbox{\boldmath{$#1$}}}

\newcommand{\beq}{\begin{equation}}
\newcommand{\eeq}{\end{equation}}
\newcommand{\beqa}{\begin{eqnarray}}
\newcommand{\eeqa}{\end{eqnarray}}
\newcommand{\beqan}{\begin{eqnarray*}}
\newcommand{\eeqan}{\end{eqnarray*}}

\begin{document}
%
%
%
\title{Temporally multiplexed superposition states of continuous variables}
%
%
%
%
%
%
\author{Masahide Sasaki}
\affiliation{
    National Institute of Information and Communications 
    Technology, 
    4-2-1 Nukui-Kita, Koganei, Tokyo 184-8795, Japan}
\affiliation{
    CREST, Japan Science and Technology Agency, 
    5 Sanbanchuoh, Chiyoda-ku, Tokyo 102-0075, Japan}

\author{Masahiro Takeoka}
\affiliation{
    National Institute of Information and Communications 
    Technology, 
    4-2-1 Nukui-Kita, Koganei, Tokyo 184-8795, Japan}
\affiliation{
    CREST, Japan Science and Technology Agency, 
    5 Sanbanchuoh, Chiyoda-ku, Tokyo 102-0075, Japan}

\author{Hiroki Takahashi}
\affiliation{
    National Institute of Information and Communications 
    Technology, 
    4-2-1 Nukui-Kita, Koganei, Tokyo 184-8795, Japan}
\affiliation{
    Department of Applied Physics, The University of Tokyo 
    7-3-1 Hongo, Bunkyo-ku, Tokyo 113-8656, Japan}
\affiliation{
    CREST, Japan Science and Technology Agency, 
    5 Sanbanchuoh, Chiyoda-ku, Tokyo 102-0075, Japan}
%
%
%
%
\date{\today} 
%
%
%
\begin{abstract}
We study non-Gaussian states generated 
by two-photon subtraction from a cw squeezed light source. 
In a cw scheme one can subtract two photons from the source 
with a designated time separation 
and can genarate temporally multiplexed 
superposition states of continuous variables.  
We numerically study the properties of these states 
in the light of bosonic interference in the time domain.  
In an appropriate temporal mode 
amplified kittens are produced in a region 
where the time separation is comparable with 
the correlation time of squeezed packets. 
\end{abstract}
%
%
%
%
\maketitle

\setcounter{section}{0}

\section{Introduction}

In optical quantum information processing 
there have been two kinds of approaches 
with discrete-variable (DV) and continuous-variable (CV) 
schemes. 
The former usually consists of entangled photon states and 
photon counters, 
while the latter consists of squeezed states and 
homodyne detectors. 
Each of these schemes is, however, only a part of control 
with full potential aspects of quantum field of light. 
The scheme combining these two will surely be promissing 
for exploiting universal quantum information processing 
for computing, communications, and metrology.

Recent progress in this direction has been made as 
the generation of 
quantum superposition of mesoscopically distinguishable states 
using a squeezed state, photon counting, and homodyne detection\,%
\cite{Ourjoumtsev_etal2006_Science,Neergaard-Nielsen_etal2006_PRL,Wakui_etal2007_OptExp,Ourjoumtsev_etal2006_PRL,Ourjoumtsev_etal2007_PRL}. 
These experiments are based on the photon subtraction 
from a squeezed state, as proposed by Dakna et al.\,%
\cite{DAOKW1997_PRA55}.  
In the scheme a small fraction of squeezed light is tapped 
via a beam splitter, and is guided into photon counters. 
The remaining light beam conditioned on photon clicks 
turns to be a highly nonclassical and non-Gaussian state, 
which was transformed from a squeezed state (Gaussian state) 
via a strong nonlinear process of photon counting. 
In this way 
quantum superposition of mesoscopically distinguishable states 
(Schr\"{o}dinger kitten states) 
were generated by the single-photon subtraction.

Next step is to produce two or more kitten states 
as well as to increase the size of cat states 
for realizing 
quantum information processing  
and
quantum metrology 
\cite{RGMMG_PRA68_2003,vanEnk_PRA64_2001,Gilchrist_JOB_QSOpt2004}. 
If larger number of photons could be subtracted from 
a squeezed light, 
larger size of quantum superpositions 
(Schr\"{o}dinger cat states) could be produced, 
although experiments get more challenging 
\cite{Nielsen_PRA76_2007}. 
Another method is to produce a larger cat state 
from two smaller cat states as proposed in 
\cite{Lund_etal2004_PRA70_020101}. 
A new method to produce larger cat states 
was recently demonstrated in laboratory, 
by using two-photon Fock state 
with conditional homodyning\,%
\cite{Ourjoumtsev_etal2007_Nature}.

In this paper, 
we study another avenue in this direction, 
which is the generation of two kitten states 
with a designated time separation in the temporal domain 
by using the two-photon subtraction 
from a cw squeezed light. 
Two kitten states are generated in a single spatial mode.  
They are of temporally two-mode, and generally entangled, 
referred to as temporally multiplexed kitten states. 
We numerically study the properties of these states 
in the light of bosonic interference in the time domain, 
and show that 
in an appropriate temporal mode 
amplified kittens are produced in a region 
where the time separation is comparable with 
the correlation time of squeezed packets. 
The paper is organized as follows. 
In Sec. \ref{rough sketch of physics}  
we introduce basic notions for our cw scheme, 
and provide a rough sketch of physics 
with an idealized lossless model. 
In Sec. \ref{precise modeling} 
we present a precise modeling in a practical setting, 
show numerical results, 
and discuss bosonic interference 
occuring in the cross over region 
where the time separation is comparable with 
the correlation time of squeezed packets.  
Section \ref{Summary} concludes the paper.

\section{Photon subtraction in a cw scheme: 
rough sketch of physics} 
\label{rough sketch of physics}

\begin{figure}
\begin{center}
\includegraphics[width=8cm]
{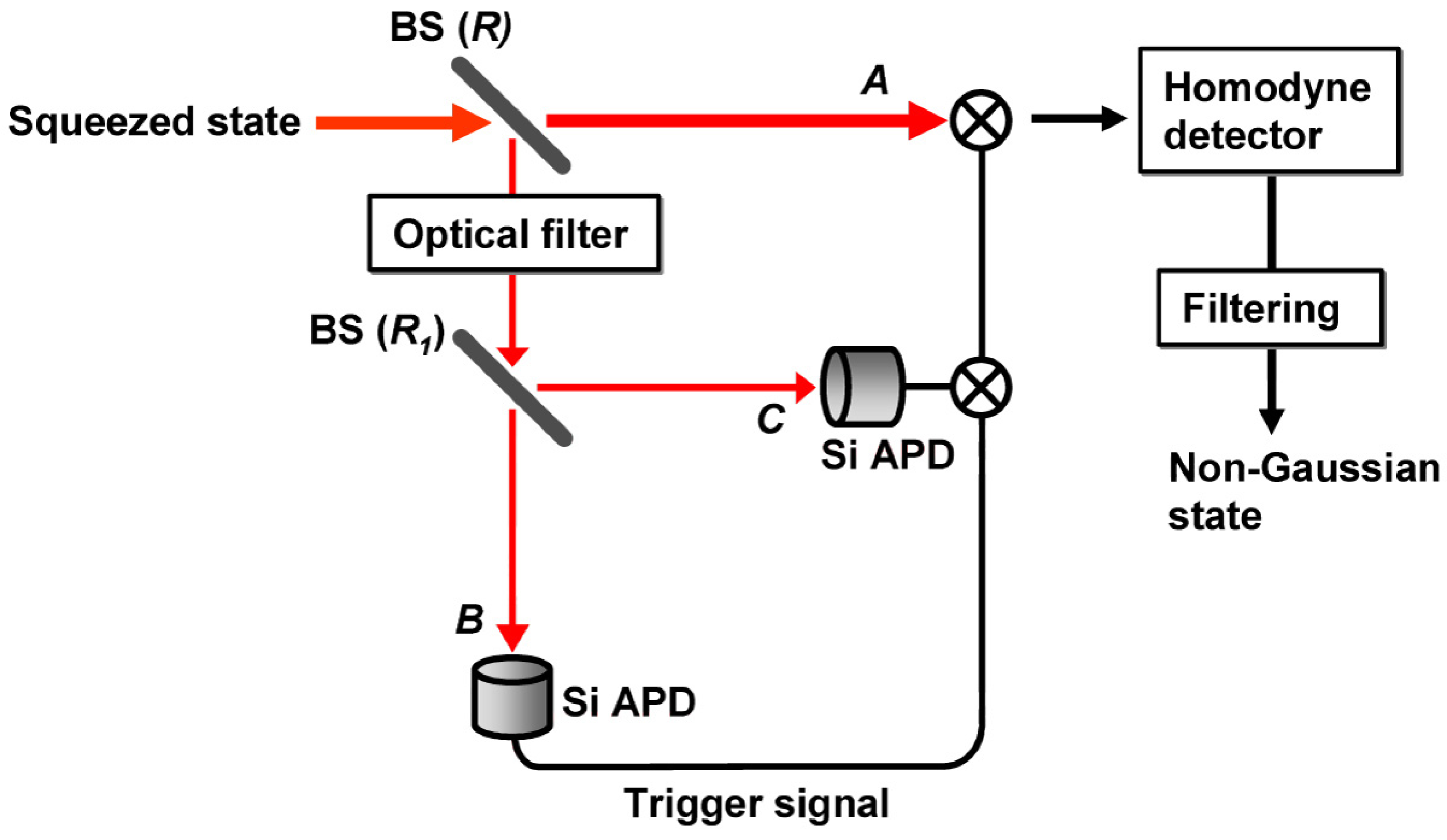}   %
\caption{\label{Photon_Subtration_Scheme}
(Color online) 
Scheme of the two-photon subtraction from a squeezed state. 
BSs are beam splitters with the reflectances of $R$ and $R_1$. 
}
\end{center}
\end{figure}

The photon subtraction scheme is depicted in Fig.
~\ref{Photon_Subtration_Scheme}. 
A small fraction of the squeezed beam 
in path A is tapped at a beam splitter (BS) 
with reflectance $R$, 
guided into two photodetectors in paths B and C 
through an optical filter and 
another beam splitter with reflectance $R_1$,  
and is then used as trigger signals 
for conditional photon subtraction.

We consider the scheme with a cw squeezed light 
generated by an optical parametric oscillator (OPO), 
as studied extensively by Nielsen and M{\o}lmer\,%
\cite{Molmer2006_PRA73_063804,Nielsen_Molmer2007_PRA75_023806,Nielsen_Molmer2007_PRA75_043801}. 
A typical feature of this scheme is that 
the trigger photodetection takes place 
in much shorter time scale ($<$1\,ns for Si APD) 
compared with the OPO time scale ($\sim$100\,ns), 
and hence interesting multimode kitten states 
can be generated in the time domain\,%
\cite{Neergaard-Nielsen_etal2006_PRL,Wakui_etal2007_OptExp}. 
This is in sharp contrast to the short pulse scheme 
where the photon counting time scale is 
longer than the pulse width\,%
\cite{Ourjoumtsev_etal2006_Science,Ourjoumtsev_etal2006_PRL,Ourjoumtsev_etal2007_PRL,Ourjoumtsev_etal2007_Nature}.

The annihilation field operator in a cw scheme 
is denoted by 
\beq  
\hat a(t)=\frac{1}{2\pi}\int_{-\infty}^{\infty} d\Omega
\hat a(\omega_0+\Omega)
e^{-i(\omega_0+\Omega)t}. 
\eeq
where 
$\omega_0$ is the center angular frequency of 
the spectrum of a laser source. 
The operator obeys the continuum commutation relation 
\beq\label{continuum commutation relation in frequency} 
[\hat a(\omega_0+\Omega), \hat a^\dagger(\omega_0+\Omega')]
=2\pi\delta(\Omega-\Omega'). 
\eeq
The time dependent field operator $\hat a(t)$ is defined in 
the interval $(-\infty,\infty)$, 
and obeys the commutation relation 
\beq\label{continuum commutation relation in time} 
[\hat a(t), \hat a^\dagger(t')]=\delta(t-t'). 
\eeq 
Squeezed light fields can be conveniently described  
in the rotating frame about 
the center frequency $\omega_0$, 
\beq\label{operator in rotating frame}
\hat A(t)=\hat a(t) e^{i\omega_0t}
=
\frac{1}{2\pi}\int_{-\infty}^{\infty} d\Omega
\hat A(\Omega) e^{-i\Omega t}, 
\eeq
where 
$\hat A(\Omega)=\hat a(\omega_0+\Omega)$.

Now let us roughly sketch an essential physics 
by assuming a nearly lossless OPO squeezing 
described by the Bogolubov transformation 
\beq\label{Bogolubov transformation by S_{A}}
\hat S_{A}^\dagger \hat A(\Omega) \hat S_{A}
  =\mu(\Omega) \hat A(\Omega)
  +\nu(\Omega) \hat A^\dagger(-\Omega). 
\eeq
The cw beam described by a pure state 
$\hat S_{A} \ket{\mathbf{0}_{A}}$ 
is the input squeezed state for photon subtraction.

\subsection{Odd-number Schr\"{o}dinger kitten 
by single-photon subtraction} 

Let us first consider the single-photon subtraction 
with a trigger signal detected at $t=t_1$ 
in the detector in path B 
($R_1=0$ in Fig.~\ref{Photon_Subtration_Scheme}). 
We assume that this detection takes place 
instantaneously compared with the OPO time scale, 
and 
the trigger beam is projected onto the vacua for other times. 
Denoting the BS operation with the reflectance $R$ 
as $\hat V_{AB}$, 
the conditional (not normalized) state is then described by 
\beqa\label{Ideal single photon subtracted state}
\ket{\rho_{CW}^{(1)}}
&\propto&
\bra{\mathbf{0}_{B}} \hat B(t_1) 
\hat V_{AB} \hat S_{A} \ket{\mathbf{0}_{AB}}
\nonumber\\
&=&
-\sqrt{R} \hat A(t_1) 
\hat S_{A} \ket{\mathbf{0}_{A}}
\nonumber\\
&=&
-\sqrt{R} \hat S_{A} \int_{-\infty}^{\infty} dt
\hat A^\dagger(t) \nu(t-t_1) \ket{\mathbf{0}_{A}}
\eeqa
where 
we have used the relation (\ref{Bogolubov transformation by S_{A}}), 
and introduced 
\beq
\nu(t-t_1)
\equiv
\frac{1}{2\pi} \int_{-\infty}^{\infty} d\Omega
e^{i\Omega (t-t_1)}
\nu(\Omega). 
\eeq
We have also assumed that $R$ is small so that 
the renormalization of the effective squeezing 
on beam A due to optical loss is negligible.

This quantity represents the temporal correlation of 
squeezed photons. 
It is roughly given by 
\beq
\nu(t) 
\sim
\frac{\epsilon}{\sqrt{\zeta_0}} \psi (t)
\eeq
where 
\beq\label{psi(t)}
\psi (t)=\sqrt{\zeta_0} e^{-\zeta_0 \vert t\vert} 
\eeq
is a normalized temporal mode function\,%
\cite{Molmer2006_PRA73_063804,Nielsen_Molmer2007_PRA75_023806,Nielsen_Molmer2007_PRA75_043801}. 
Here $\zeta_0/\pi$ corresponds to the OPO resonant bandwidth, 
and $\epsilon$ is the nonlinear coefficient of the OPO, 
propotional to the $\chi^{(2)}$ interaction coefficient 
and the pump field amplitude. 
Defining the field operator localized in this temporal mode by 
\beq\label{A_1}
\hat A_1^\dagger=\int_{-\infty}^{\infty} dt
\hat A^\dagger(t) \psi (t-t_1) 
\eeq
the single-photon subtracted state can be written as 
\beq\label{Ideal single photon subtracted state2}
\ket{\rho_{CW}^{(1)}}
\propto
-\frac{{\sqrt{R}}\epsilon}{\sqrt{\zeta_0}} 
\hat S_{A} \hat A_1^\dagger \ket{\mathbf{0}_{A}}. 
\eeq
This represents a cw beam consisting of 
the squeezed single-photon state\,%
\cite{Lund_etal2004_PRA70_020101,Suzuki_etal2006_OCom259_758} 
localized in the mode $\psi (t-t_1)$ 
and the squeezed vacua in the other time domains.

The beam is filtered through the mode function $\psi (t-t_1)$, 
and then sampled by homodyne detection\,%
\cite{Molmer2006_PRA73_063804}. 
In the homodyne output one can see that 
two distinguishable waveforms with phase difference of $\pi$ 
coexist as demonstrated experimentally in\,%
\cite{Ourjoumtsev_etal2006_Science,Neergaard-Nielsen_etal2006_PRL,Wakui_etal2007_OptExp}. 
The constructed Wigner functions have negative part, 
showing the nonclassicality of the states. 
This is a quantum superposition of mesoscopically distinguishable 
states as analyzed by Dakna et al. 
\cite{DAOKW1997_PRA55}, 
the so called {\it Schr\"{o}dinger kitten states}.   
In this case it is an odd-number kittens.

\subsection{Entangled kittens by two-photon subtraction} 

In the two-photon subtraction, 
we set $R_1=0.5$ (i.e. 50\%) in Fig.
~\ref{Photon_Subtration_Scheme}. 
Suppose that a first and second photon are detected 
at $t=t_1$ in path B, 
and 
at $t=t_2$ in path C, respectively. 
The conditional state can be represented as 
\beqa\label{Ideal two photon subtracted state}
&&\ket{\rho_{CW}^{(2)}}
\nonumber\\
&&\propto
\bra{\mathbf{0}_{BC}}
    \hat C(t_2)\hat B(t_1)\hat V_{BC}\hat V_{AB} \hat S_A 
  \ket{\mathbf{0}_{ABC}}
\nonumber\\
&&=
\frac{R}{2}
\hat A(t_2) \hat A(t_1) 
\hat S_{A} \ket{\mathbf{0}_{A}}
\nonumber\\
&&=
\frac{R}{2} \hat S_{A} 
\Biggl[
\int_{-\infty}^{\infty} dt'
\hat A^\dagger(t') \nu(t'-t_2)
\int_{-\infty}^{\infty} dt
\hat A^\dagger(t) \nu(t-t_1)
\nonumber\\
&&\qquad\quad
+
\int_{-\infty}^{\infty} \frac{d\Omega}{2\pi}
e^{i\Omega (t_2-t_1)} \mu(-\Omega)\nu(\Omega) 
\Biggr]
\ket{\mathbf{0}_{A}}
\nonumber\\
&&
\sim
\frac{R}{2} \hat S_{A} 
\biggl[
   \frac{\epsilon^2}{\zeta_0}
   \hat A_2^\dagger \hat A_1^\dagger
+\frac{\epsilon}{\sqrt\zeta_0}\psi (t_2-t_1)
\biggr]
    \ket{\mathbf{0}_{A}}. 
\eeqa
One can immediately see the two extremes;

(i) $\Delta$\,$(\equiv\vert t_2 - t_1 \vert)\ll\zeta_0^{-1}$ 
where 
the two-photon subtraction takes place in the single mode,  
producing an even-number kitten as originally proposed in 
\,%
\cite{DAOKW1997_PRA55}. 
%

(ii) $\Delta\gg\zeta_0^{-1}$ where 
two odd-number kittens are generated in separated packets 
$\psi(t-t_1)$ and $\psi(t-t_2)$. 
Each kitten is the one generated in
\,%
\cite{Neergaard-Nielsen_etal2006_PRL,Wakui_etal2007_OptExp}.

In the imtermediate region there is a cross over 
between (i) and (ii), 
and both even- and odd-kittens exsist, 
being entangled over the two modes. 
This region is of our interest.

Since the field operators $\hat A_1^\dagger$ and $\hat A_2^\dagger$ 
have generally a finite mode overlap for finite $\Delta$
\beqa 
 I_\Delta
&\equiv&
\bra{\mathbf{0}_{A}}\hat A_2 
\hat A_1^\dagger \ket{\mathbf{0}_{A}}
\nonumber
\\
&=&
\int_{-\infty}^{\infty} dt
\psi(t-t_1)\psi(t-t_2)
\nonumber
\\
&=&
( 1+\zeta_0\Delta )e^{-\zeta_0\Delta} 
\eeqa
one needs an orthonormal mode set to describe the state.

\subsubsection{Unbiased modes} 

\begin{figure}
\begin{center}
\includegraphics[width=8cm]
{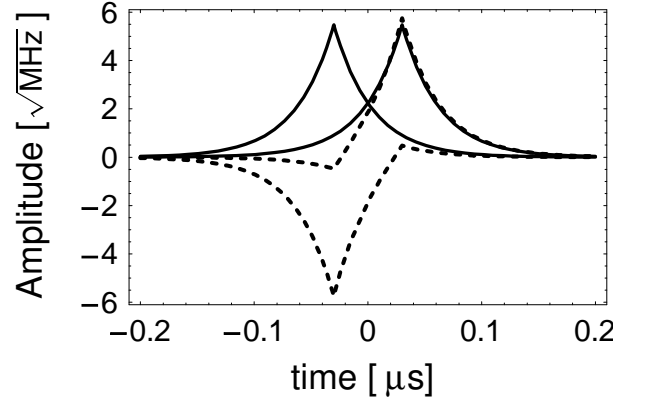}   %
\caption{\label{Unbiased_mode_funcs}
Temporal shape of the unbiased modes (dashed lines) 
in the case of $\zeta_0=30$\,MHz. 
The solid lines are $\psi(t-t_1)$ and $\psi(t-t_2)$. 
}
\end{center}
\end{figure}

\begin{figure}
\begin{center}
\includegraphics[width=6cm]
{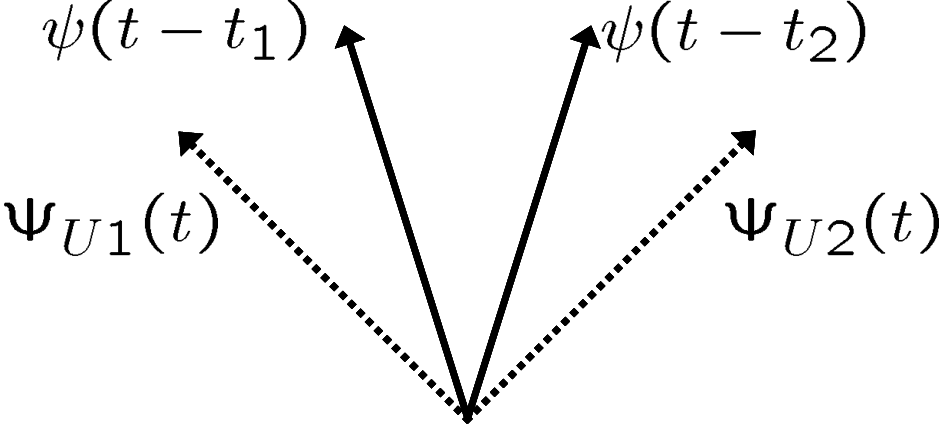}   %
\caption{\label{Unbiased_modes}
Vector representation of the unbiased modes. 
}
\end{center}
\end{figure}

We first consider an orthonormal mode function pair 
$\{\Psi_{U1}(t),\Psi_{U2}(t)\}$ 
which suits to analyze the region $\zeta_0\Delta>1$, 
and defined by 
\beqa\label{Unbiased modes}
\psi(t-t_1)&=&
\sqrt{1-p_\Delta}\Psi_{U1}(t)+\sqrt{p_\Delta}\Psi_{U2}(t)
\nonumber\\
\psi(t-t_2)&=&
\sqrt{p_\Delta}\Psi_{U1}(t)+\sqrt{1-p_\Delta}\Psi_{U2}(t)
\eeqa
where 
\beq 
p_\Delta=\frac{1-\sqrt{1- I_\Delta^2}}{2}. 
\eeq
We call them {\it the unbiased modes} because 
$\Psi_{U1}(t)$ and $\Psi_{U2}(t)$ equally overlap with 
$\psi(t-t_1)$ and $\psi(t-t_2)$, respectively.  
Their temporal shapes are shown in Fig.~\ref{Unbiased_mode_funcs} 
as well as 
the mode overlap configuration in vector representation 
in Fig.~\ref{Unbiased_modes}. 
The corresponding field operators are defined by 
\beq\label{Unbiased field operators}
\hat A_{Ui}\equiv
\int_{-\infty}^{\infty} dt
\hat A(t) \Psi_{Ui}(t)
\eeq
satisfying 
\beq
[\hat A_{Ui}, \hat A_{Uj}^\dagger]=\delta_{ij}. 
\eeq
The operators $\hat A_i^\dagger$ are represented as  
\beqa\label{Unbiased field operators:relation}
\hat A_1^\dagger&=&
   \sqrt{1-p_\Delta}\hat A_{U1}^\dagger
  +\sqrt{p_\Delta}\hat A_{U2}^\dagger
\nonumber\\
\hat A_2^\dagger&=&
   \sqrt{p_\Delta}\hat A_{U1}^\dagger
  +\sqrt{1-p_\Delta}\hat A_{U2}^\dagger. 
\eeqa
The two-photon subtracted state (not noralized) is 
then represented as 
\beqa\label{Two photon subtracted state in unbiased modes}
&&
\hat A(t_2) \hat A(t_1) 
\hat S_{A} \ket{\mathbf{0}_{A}}
\nonumber\\
&&
\sim
\hat S_{A} 
\Biggl[
\frac{\epsilon^2}{\zeta_0}
\biggl(\ket{1,1}
   +
    I_\Delta
   \frac{\ket{2,0}+\ket{0,2}}{\sqrt2}
\biggr)
\nonumber\\
&&
\qquad\quad  +\epsilon e^{-\zeta_0\Delta}\ket{0,0}
\Biggr]
\ket{\mathbf{0}_{\tilde A}}. 
\eeqa
The ket vectors in the above equation  
are understood as 
$\ket{\cdots,\cdots}=\ket{\cdots}_{U1}\otimes\ket{\cdots}_{U2}$. 
Thus the state in terms of the unbiased modes is invariant 
under the mode permutation. 
The $\Delta$-dependence of the above state 
is completely governed by the coefficients in the bracket. 
Note that $\hat S_A$ is independent of $\Delta$.  
The index $\tilde A$ means the other remaining modes than 
those spanned by the $U1$ and the $U2$. 
The modes $\tilde A$ will be filtered out 
to observe the two-photon subtracted states.

The two-mode state is essentially an entangled state 
of the two kinds of kittens 
(the odd-number and even-number kittens), 
and the squeezed vacuum. 
In fact the two-photon subtracted state can also be represented as 
\beqa\label{Two photon subtracted state in unbiased modes 2}
\ket{\rho_{CW}^{(2)}}
&\equiv&
\frac{\hat A(t_2) \hat A(t_1) \hat S_{A} \ket{\mathbf{0}_{A}}}
{||\hat A(t_2) \hat A(t_1) \hat S_{A} \ket{\mathbf{0}_{A}}||}
\nonumber\\
&\sim&
{\cal N}\hat S_{A} 
\Biggl[
\frac{\epsilon}{\zeta_0}\ket{1,1}
\nonumber\\
&&
+e^{-\zeta_0\Delta}\nu_e
   \biggl( \ket{\phi_e,0}+\ket{0,\phi_e} \biggr)
\Biggr]
\ket{\mathbf{0}_{\tilde A}}.  
\eeqa
where
\beqa\label{phi_e etc}
\ket{\phi_e}&\equiv&\nu_e^{-1}
\Biggl[
\frac{\epsilon}{\zeta_0}(1+\zeta_0\Delta)\ket{2}
+\frac{1}{\sqrt2}\ket{0}
\Biggr]
\nonumber\\
{\cal N}&\equiv&
\Biggl[\frac{\epsilon^2}{\zeta_0^2}
+e^{-2\zeta_0\Delta}\biggl(\nu_e^2+\frac{1}{2}\biggr)
\Biggr]^{-1/2}
\nonumber\\
\nu_e&\equiv&
\sqrt{\frac{\epsilon^2}{\zeta_0^2}(1+\zeta_0\Delta)^2
      +\frac{1}{2}}.
\eeqa
The first term includes two odd-number kittens 
in the similar sense of 
Eq. (\ref{Ideal single photon subtracted state2}). 
As for the second term, 
when $\ket{\phi_e}$ is squeezed, 
one has an even-number kitten as demonstrated by Dakna et al. 
\cite{DAOKW1997_PRA55}. 
(Note that 
this is not precisely a coherent-state superposition 
but a superposition of other kinds of two mesoscopically 
distinct components with even-number photons. 
The detailed features will be presented in the next section 
numerically.)

The characteristic features are summarized as follows:

(i) 
The first term in Eq. 
(\ref{Two photon subtracted state in unbiased modes 2}) 
includes the two single-photon-state kernels, 
each comes from the {\it distinguishable} photon subtraction 
from each squeezed light packet 
in $\psi(t-t_1)$ and $\psi(t-t_2)$. 
This corresponds to the two odd-number kittens. 
(See Eq. (\ref{Ideal single photon subtracted state2}).)

(ii) 
The second and third terms with $\ket{\phi_e}$ in Eq. 
(\ref{Two photon subtracted state in unbiased modes 2}) 
include the two even-number-state kernels. 
They come from the {\it indistinguishable} photon subtraction 
from the overlapping components between the two localized 
packets in $\psi(t-t_1)$ and $\psi(t-t_2)$, 
and their weights decreases as $\Delta$. 
The $\ket{\phi_e}$ being squeezed, 
the resulting state is an even-number kitten.

\subsubsection{Biased modes} 

\begin{figure}
\begin{center}
\includegraphics[width=8cm]
{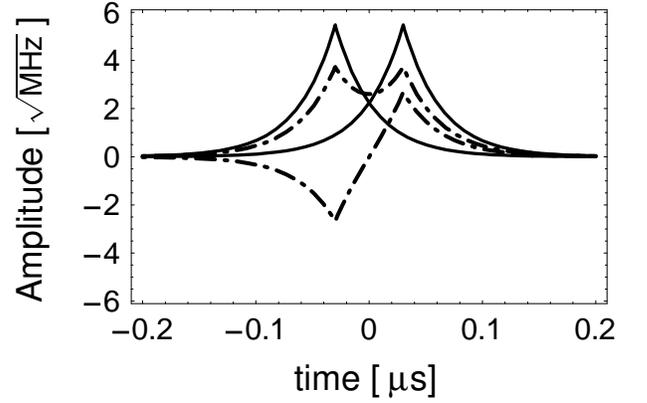}   %
\caption{\label{Biased_mode_funcs}
Temporal shape of the biased modes 
(one-dotted lines) in the case of $\zeta_0=30$\,MHz. 
The $\Psi_+(t)$ is symmetric, 
while the $\Psi_-(t)$ is asymmetric. 
The solid lines are $\psi(t-t_1)$ and $\psi(t-t_2)$. 
}
\end{center}
\end{figure}

\begin{figure}
\begin{center}
\includegraphics[width=6cm]
{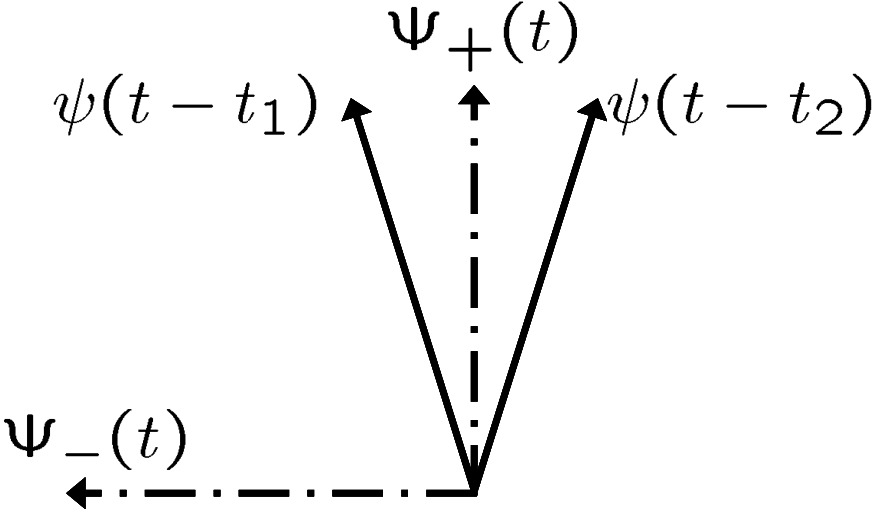}   %
\caption{\label{Biased_modes}
Vector representation of the unbiased modes. 
}
\end{center}
\end{figure}

Let us then consider the other important mode function pair, 
which suits to analyze the region $\zeta_0\Delta<1$, 
and defined by  
\beq
\Psi_\pm(t)=
\frac{\psi(t-t_2)\pm\psi(t-t_1)}
     {\sqrt{ 2\left(
              1\pm  I_\Delta \right) } }. 
\eeq
Their temporal shapes are shown in Fig.~\ref{Biased_mode_funcs} 
as well as 
the mode overlap configuration in vector representation 
in Fig.~\ref{Biased_modes} . 
The $\Psi_+(t)$ is the mode that extracts the features 
due to the overlap of the packets 
$\psi(t-t_1)$ and $\psi(t-t_2)$, 
while the $\Psi_-(t)$ is the mode that extracts 
the differential features of these localized packets. 
We refer them to as {\it the biased modes}.  
Defining the field operators in the biased modes by 
\beq\label{signal mode field}
\hat A_\pm
\equiv 
\int_{-\infty}^{\infty} dt \hat A(t) \Psi_\pm(t)  
\eeq
and the number states by 
$
\ket{n_+}
=
\frac{\hat A_+^{\dagger n}}{\sqrt{n!}}\ket{0_+}
$
%
%
%
the two-photon subtracted state can either be represented as 
\beqa\label{Two photon subtracted state in biased modes}
&&
\hat A(t_2) \hat A(t_1) 
\hat S_{A} \ket{\mathbf{0}_{A}}
\nonumber\\
&&
\sim
\hat S_{A} 
\Biggl[
\frac{\epsilon^2}{\zeta_0}
\biggl(
\frac{1+ I_\Delta}{\sqrt2}\ket{2_+,0_-}
-
\frac{1- I_\Delta}{\sqrt2}\ket{0_+,2_-}
\biggr)
\nonumber\\
&&
\qquad\quad +\epsilon e^{-\zeta_0 \Delta} 
\ket{0_+,0_-}
\Biggr]
\ket{\mathbf{0}_{\tilde A}}.
\eeqa

Thus in the biased mode 
the output state is a squeezed entangled-state 
of 2-photon and the vacuum states over modes $\Psi_\pm(t)$. 
This is actually an entangled state of 
even-number kitten state and the squeezed vacuum state. 
For $\zeta_0\Delta\ll1$ 
\beqa\label{Two photon subtracted state in biased modes 3}
\hat A(t_2) \hat A(t_1) 
\hat S_{A} \ket{\mathbf{0}_{A}}
&\rightarrow&
\hat S_{A+}
\biggl( \sqrt{2}\frac{\epsilon^2}{\zeta_0}\ket{2_+}
       +\epsilon \ket{0_+} \biggr)
\nonumber\\
&&
\otimes\hat S_{A-}\ket{0_-}
\otimes\hat S_{\tilde A} \ket{\mathbf{0}_{\tilde A}}
\eeqa
where we have approximately decompose the squeezing operator 
$\hat S_{A}$ into a product of the operators for the modes 
$\Psi_+(t)$, $\Psi_-(t)$ and others. 
Most of photons are contained in mode $\Psi_+(t)$ 
while 
mode $\Psi_-(t)$ is nearly the vacuum state. 
Actually the squeezing in the aymmetric mode $\Psi_-(t)$ 
is very small. 
This feature holds for $\zeta_0\Delta\alt1$, 
and  
for certain finite $\Delta$'s, 
an amplified even-number cat state is produced 
in mode $\Psi_+(t)$ 
as shown in the next section.

For $\zeta_0\Delta\gg1$ on the other hand 
photons are distributed equally to 
modes $\Psi_+(t)$ and $\Psi_-(t)$ as
\beqa\label{Two photon subtracted state in biased modes 4}
&&\hat A(t_2) \hat A(t_1) 
\hat S_{A} \ket{\mathbf{0}_{A}}
\nonumber\\
&&\rightarrow
\frac{\epsilon^2}{\sqrt{2}\zeta_0}
\hat S_{A+}\otimes\hat S_{A-}
\biggl(\ket{2_+,0_-}-\ket{0_+,2_-}
\biggr)
\nonumber\\
&&\quad
\otimes\hat S_{\tilde A} \ket{\mathbf{0}_{\tilde A}}
\eeqa

Thus the two-photon subtraction from a cw squeezed beam 
generates temporally multiplexed Schr\"{o}dinger kittens 
over the two temporal modes. 
In the next section we show the features described above 
by calculating the Wigner functions 
based on the precise modeling in a practical setting. 
A new formalism to treat only the two modes of interest 
with the effective super-operators and detailed analysis 
of the states are presented elswhere
\,
\cite{Takeoka2008_Large_cat}.

\section{Photon subtraction in a cw scheme: 
precise modeling} 
\label{precise modeling}

\subsection{Model} 

In this section 
we extend the idealized lossless model described above 
to a practical case, 
and provide the formulae and numerical results for 
future experiments. 
In a practical setting of OPO, 
the pure-state argument in the previous section cannot apply 
because the two temporal modes have generally 
quantum correlations with the other modes 
(other neighboring squeezed packets). 
This makes the relevant states mixed more or less. 
One needs to calculate density operators, 
whose details are given in Appendix.

A cw squeezed state generated by an OPO, 
$\hat\rho_A^{(0)}$,  
is fully characterized in the time domain 
by the correlation functions\,%
\cite{Drummond_Reid1990_PRA41_Squeezing} 
\beqa\label{time correlation}
&&\tr[\hat\rho_A^{(0)} \hat A^\dagger(t) \hat A(t') ]
\nonumber\\
&&=
\frac{\epsilon\gamma_T}{4}
\Biggl(
\frac{e^{-\zeta(-\epsilon)|t-t'|}}{\zeta(-\epsilon)}
-
\frac{e^{-\zeta( \epsilon)|t-t'|}}{\zeta( \epsilon)}
\Biggr)
\\
&&\tr[\hat\rho_A^{(0)} \hat A(t) \hat A(t') ]
\nonumber\\
&&=
\frac{\epsilon\gamma_T}{4}
\Biggl(
\frac{e^{-\zeta(-\epsilon)|t-t'|}}{\zeta(-\epsilon)}
+
\frac{e^{-\zeta( \epsilon)|t-t'|}}{\zeta( \epsilon)}
\Biggr) 
\eeqa
where 
\beq
\zeta(\epsilon)\equiv
\zeta_0+\epsilon, 
\quad
\zeta_0\equiv\frac{\gamma_T+\gamma_L}{2}
\eeq
with the leakage rates  
$\gamma_T$ of the output coupler 
and 
$\gamma_L$ of the cavity loss. 
The $\zeta_0$ 
determines the resonant bandwidth of the cavity. 
In a typical experiment with a low loss $\chi^{(2)}$-crystal\,%
\cite{Wakui_etal2007_OptExp}, 
$\gamma_T\sim57$\,MHz, $\gamma_L\sim1.2$\,MHz, 
and $\zeta_0/\pi\sim9.3$\,MHz (the full width at half maximum).

A small fraction is taken from the squeezed state 
and is split into paths B and C as the trigger beams, 
resulting the three-beam squeezed state 
\beq\label{rho_ABC}
\hat\rho_{ABC}
=
\hat V_{BC} \hat V_{AB} \hat\rho_A^{(0)} 
\hat V_{AB}^\dagger \hat V_{BC}^\dagger.   
\eeq 
The trigger beams are then measured by two on/off photodetectors 
with the time resolution $T$ ($\sim1$\,ns). 
Its mathematical modeling is described in 
\,\cite{Sasaki_Suzuki_PRA73_2006}.  
Denoting the ``on"-signal elements of 
the positive operator-valued measure (POVM) 
of the on/off detectors as 
$\bar{\Pi}_B$ and $\bar{\Pi}_C$, 
the output state in path A conditioned by the ``on" signals 
is given by 
\beq\label{rho_A by two photon subtraction}
\hat \rho_A
=
\frac{1}{P_{det}}
\tr_{BC} 
\left( \hat\rho_{ABC} \bar{\Pi}_B\otimes\bar{\Pi}_C \right) 
\eeq
where 
\beq\label{P_det}
P_{det}
=
\tr_{ABC} \left( 
           \hat\rho_{ABC} \bar{\Pi}_B\otimes\bar{\Pi}_C
           \right) 
\eeq
is the detection probability.

This state is still a cw beam, consisting of multimodes. 
We are, however, interested in the two modes; 
the unbiased modes $\{ \Psi_{U1}(t), \Psi_{U2}(t)\}$ 
and the biased modes $\{ \Psi_+(t), \Psi_-(t)\}$.  
Denoting these sets as $\{\Psi_k(t)\}$  
we introduce the annihilation operators  
\beq\label{hat A_k}
\hat A_k
\equiv 
\int_{-\infty}^{\infty} dt \hat A(t) \Psi_k(t).  
\eeq
Let the quadrature amplitude and phase operators be 
\beq\label{quadrature for beam A}
\hat X_k
=
\frac{\hat A_k + \hat A_k^{\dagger}}{\sqrt2}, 
\quad 
\hat P_k
=
\frac{\hat A_k - \hat A_k^{\dagger}}{{\sqrt2}i}. 
\eeq
In the homodyne channel 
one can observe a reduced state from $\hat \rho_A$ 
into a desired mode by filtering the homodyne current 
into 
\beq\label{quadrature phi for beam A}
\hat X_k(\phi)
=
\frac{  \hat A_k e^{-i\phi} 
      + \hat A_k^{\dagger}e^{i\phi}}{\sqrt2} 
\eeq
and by constructing the Wigner functions 
with respect to this quadrature amplitude.

\subsection{Numerical results} 

\begin{figure}
\hspace{10mm}
\begin{center}
\includegraphics[width=8cm]
{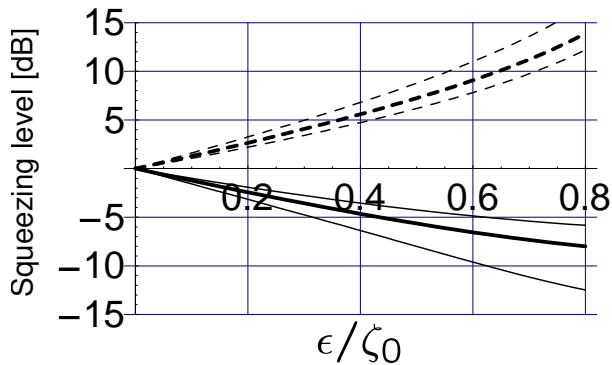}   %
\caption{\label{Sq_z}
Squeezing and antisqueezing characteristics 
of the tempral mode $\psi(t)={\sqrt f}e^{-f |t|}$ 
as a function of $\epsilon/\zeta_0$. 
The solid (dashed) lines are for 
$f=10$, 30, and 50\,MHz, from the bottom (top). 
}
\end{center}
\end{figure}

\begin{figure*}
\begin{center}
\includegraphics[width=0.9\linewidth]
{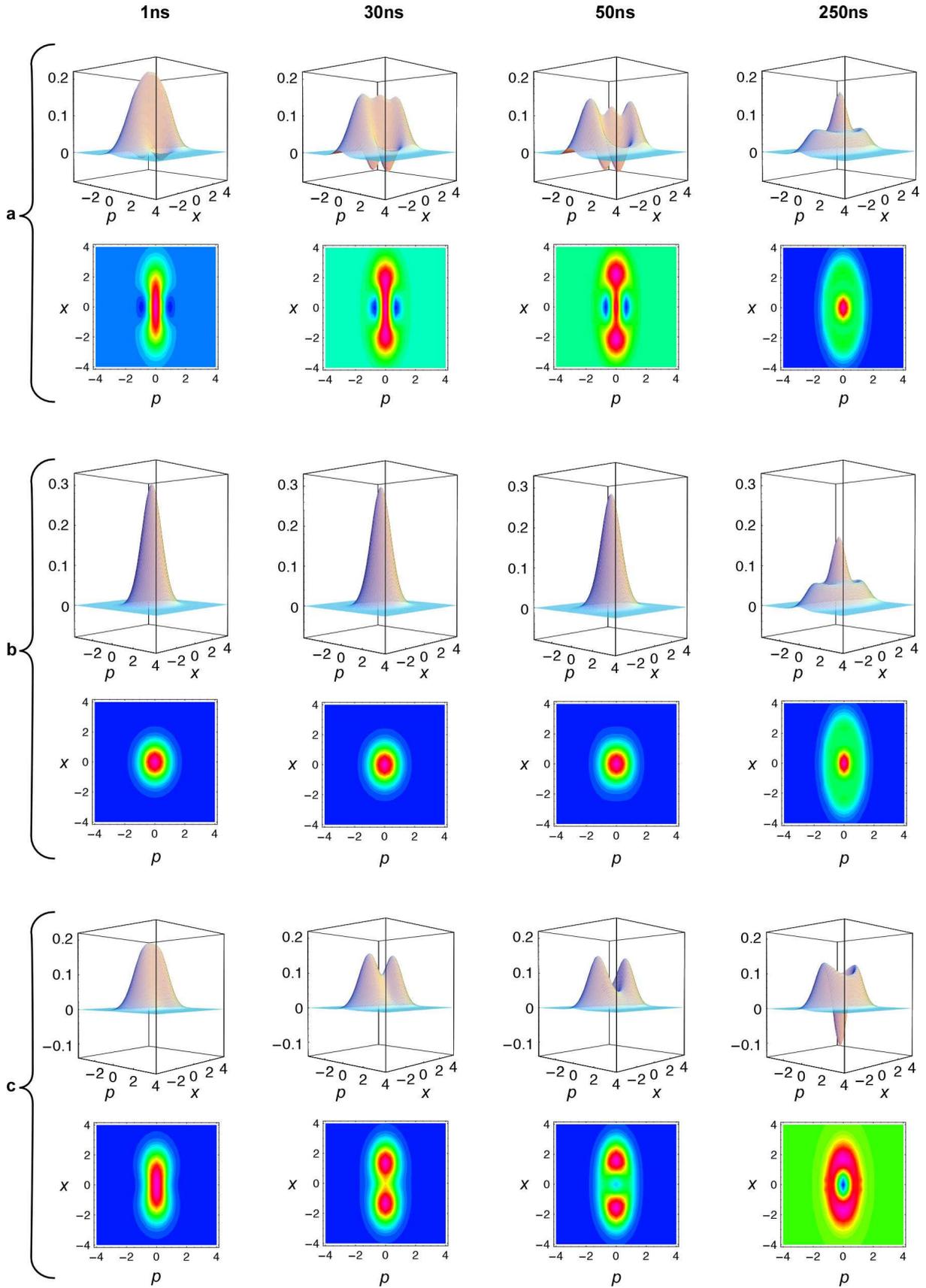}   %
\caption{\label{Wig_D_dep_all}
Wigner functions and their contour plots. 
a for the biased mode $\Psi_+(t)$. 
b for the biased mode $\Psi_-(t)$. 
c for the unbiased mode $\Psi_{U1}(t)$. 
From left to right $\Delta=1$, 30, 50, and 250\,ns. 
}
\end{center}
\end{figure*}

In Fig.~\ref{Sq_z} the squeezing and antisqueezing 
characteristics of the input beam are shown 
for the tempral mode 
$\psi(t)={\sqrt f}e^{-f |t|}$ 
for $f=10$, 30, and 50\,MHz. 
Since the squeezing itself is larger at lower frequencies, 
filtering with wider tempral shape observes 
larger squeezing.  
But the photon-subtracted squeezed state in the cw scheme 
is most appropriately observed for 
a characteristic frequency of the OPO 
$f\sim\zeta_0\sim30$MHz 
(the thick solid and dashed lines in Fig.~\ref{Sq_z}). 
The reason for it includes 
a fact that the photon counting by Si APD 
takes place in a time scale much shorter than 
the squeezing correlation time.

Figure~\ref{Wig_D_dep_all} shows the Wigner functions 
(see Appendix for the formulae) 
and their contour plots; 
a is for the biased mode $\Psi_+(t)$, 
b is for the biased mode $\Psi_-(t)$,  
and c is for the unbiased mode $\Psi_{U1}(t)$. 
The nonlinear coefficient is taken as $\epsilon/\zeta_0=0.3$, 
corresponding to -3.6 dB squeezing for the localized mode 
$\psi(t)={\sqrt \zeta_0}e^{-\zeta_0 |t|}$. 
The reflectance of the tapping BS is set to $R=0.05$ (i.e. 5\%). 
The quantum efficiency of the homodyne detector is taken as 
$\eta_H=0.96$ in Eq. (\ref{eta_H}). 
For the on/off detectors, 
the overall detection efficiency of $\eta=0.6$ and 
the fake trigger rate of $\nu=10^{-7}$ 
are assumed 
in Eq. (\ref{Characteristic func of bar Pi_B:def}). 
Since the Wigner functions are identical for both of the 
unbiased modes, only the ones for $\Psi_{U1}(t)$ are shown. 
From left to right $\Delta=1$, 30, 50, and 250\,ns. 
Here note that 
{\it the original two-mode state 
is essentially an entangled state}. 
So a reduced state to a particular mode is more or less 
a mixed state. 
But one can see typical features of kitten components behind.

The left two columns are typical cases of $\zeta_0\Delta\alt1$.  
For $\Delta=1$\,ns an even-number kitten is 
generated in mode $\Psi_+(t)$, 
while a nearly vacuum state is in mode $\Psi_-(t)$. 
The reduced state to mode $\Psi_{U1}(t)$ seems close to 
a thermalized squeezed vacuum. 
For $\Delta=30$\,ns an amplified even-number kitten 
appears in mode $\Psi_+(t)$ with a deeper negative 
Wigner function. 
The reduced state to mode $\Psi_{U1}(t)$ 
becomes close to a mixture of two Gaussian states.

The right two columns are typical cases of $\zeta_0\Delta>1$. 
For $\Delta=50$\,ns 
the reduced state to mode $\Psi_+(t)$ 
still has a negative Wigner function, and larger amplitude.   
The state seen in mode $\Psi_{U1}(t)$ shows 
a dip at the phase space origin, 
which is a typical feature of odd-number kitten.  
Actually $W(0,0)$ directly reflects the parity of 
dominant photon number, namely a dip for odd and peak for even. 
For $\Delta=250$\,ns 
one has two odd-number kittens as seen in mode $\Psi_{U1}(t)$. 
In terms of the biased modes $\Psi_\pm(t)$ 
the states reduces to identical mixed states of 
the even-number cat and the squeezed vacuum.

\begin{figure*}
\begin{center}
\includegraphics[width=0.88\linewidth]
{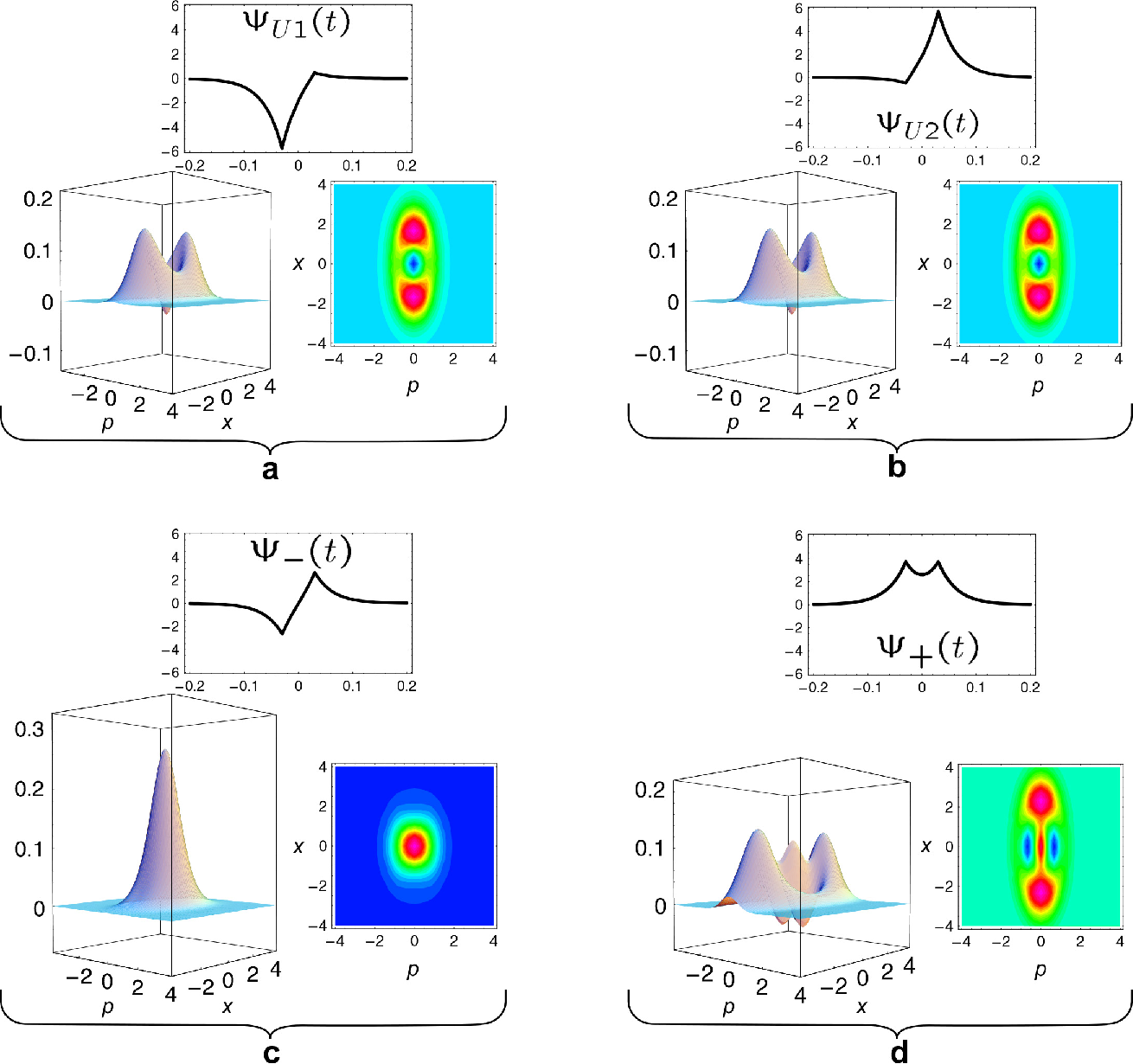}   %
\caption{\label{Wig_UB_B}
Wigner functions and their contour plots 
for the two sets of modes 
in the case of $\Delta=65$\,ns. 
}
\end{center}
\end{figure*}

Figure~\ref{Wig_UB_B} shows the case of $\Delta=65$\,ns. 
The upper panels a and b show 
the Wigner functions and their contour plots 
of the states reduced to the unbiased modes, 
while 
the lower panels c and d show the ones 
of the states reduced to the biased modes. 
For this time separation 
the reduced states in both kinds of modes 
show the negative Wigner functions. 
The panels a and b correspond to the odd-number kittens, 
while 
the pannel c and d correspond to the squeezed vacuum 
and the even-number cat. 
(These reduced states are more or less mixed ones 
due to the two-mode entanglement behind.) 
Thus by changing the filtering functions 
one can access different features of the components. 
One possible way of viewing the change from 
a and b to c and d in Fig.~\ref{Wig_UB_B} is 
a constructive bosonic interference of 
the mesoscopic superposition states 
in the sense that 
odd-number-rich components seen in modes 
$\Psi_{U1}(t)$ and $\Psi_{U2}(t)$
condensate into mode $\Psi_+(t)$ constructively, 
producing an amplified even-number-rich cat. 
On the other hand, 
in mode $\Psi_-(t)$ the destructive interference 
occurs, leaving the vacuum-like state.

\begin{figure*}
\begin{center}
\includegraphics[width=0.8\linewidth]
{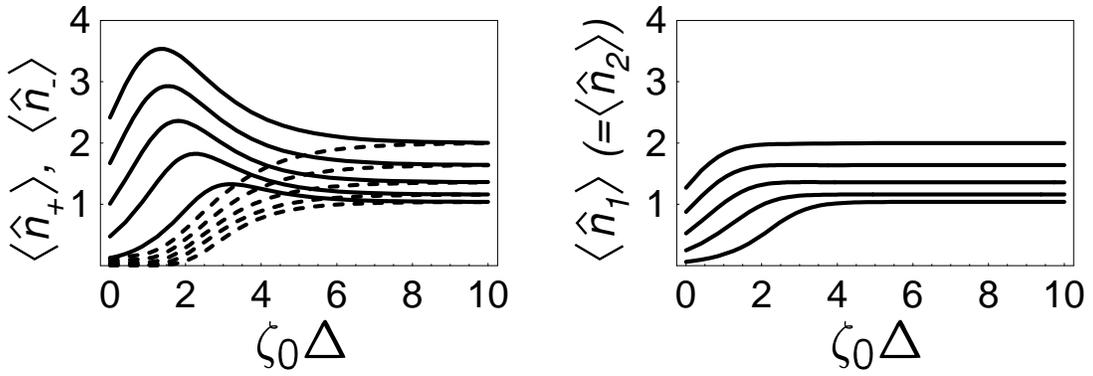}   %
\caption{\label{n_D_dep}
Average photon numbers as a function of $\zeta_0\Delta$ 
for five kinds of squeezing levels, 
from the bottom $\epsilon/\zeta_0=0.1$, 
0.2, 0.3, 0.4 and 0.5. 
The left is for the biased modes $\Psi_+(t)$ (solid line) 
and $\Psi_-(t)$ (dashed line).  
The right is for one of the unbiased modes. 
}
\end{center}
\end{figure*}

In Fig.~\ref{n_D_dep} we depict 
average photon numbers as a function of $\Delta$ 
for five kinds of squeezing levels, 
from the bottom $\epsilon/\zeta_0=0.1$ (-1.2dB), 
0.2 (-2.4dB), 0.3 (-3.6dB), 0.4 (-4.7dB) and 0.5 (-5.7dB). 
See Fig.~\ref{Sq_z} for the corresponding squeezing levels 
for $f=30$\,MHz. 
(The formulae are given in the last part of Appendix.) 
The left is for the biased modes $\Psi_+(t)$ (solid line) 
and $\Psi_-(t)$ (dashed line).  
The right is for one of the unbiased modes. 
Since $\langle \hat n_1\rangle=\langle \hat n_2\rangle$ 
for the unbiased modes, 
the right panel directly show how the total average photon 
number increases and saturates to a certain level 
as $\zeta_0\Delta$ increases.

\begin{figure}
\begin{center}
\includegraphics[width=8cm]
{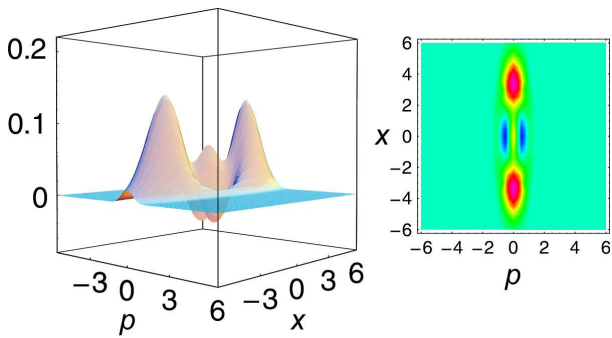}   %
\caption{\label{WigCntrP43ns5z}
Wigner function and its contour plot 
of the state in mode $\Psi_+(t)$ 
around the peak of $\langle \hat n_+\rangle$, 
corresponding to 
$\epsilon/\zeta_0=0.5$ and $\Delta=43$\,ns. 
}
\end{center}
\end{figure}

An interesting feature is the peak structure 
in mode $\Psi_+(t)$ seen in the left panel. 
Most of photons accumulate in mode $\Psi_+(t)$ 
around $\zeta_0\Delta\sim1\sim3$, 
depending on the squeezing levels. 
On the other hand, the average photon number 
in mode $\Psi_-(t)$ 
is suppressed in this time range. 
Roughly speaking amplified even-number kittens 
appear in this time range. 
A typical Wigner function for larger squeezing 
$\epsilon/\zeta_0=0.5$ with $\Delta=43$\,ns 
is shown in Fig.~\ref{WigCntrP43ns5z}. 
This is close to a quantum superposition of 
distinct squeezed states with 
phase difference of $\pi$. 
Interestingly this is quite similar 
to the cat state generated by four-photon subtraction, 
which was predicted in 
\cite{DAOKW1997_PRA55}. 
Precisely speaking 
the maximum quantumness of the cats 
including the negativity of the Wigner function 
and the state purity does not necessarily 
coincide to the peak of $\langle \hat n_+\rangle$. 
The detailed analysis will be presented 
in another paper 
\cite{Takeoka2008_Large_cat}.

The process of producing the amplified kittens 
is nothing but the constructive bosonic interefence 
caused by nonlinear process of photon subtraction. 
This is directly seen as the superposition in 
Eq. (\ref{Two photon subtracted state in biased modes}). 
The mechanism behind this consists of the following two aspects. 
The first one is the indistingushable two-photon subtraction 
from the temporally overlapped squeezed packets. 
This aspect is represented by the term 
$\hat S_{A}\hat A_2^\dagger \hat A_1^\dagger\ket{\mathbf{0}_{A}}$ 
in Eq. (\ref{Ideal two photon subtracted state}), 
or equivalently the first and second terms in 
Eq. (\ref{Two photon subtracted state in biased modes}). 
Note that the two-photon subtraction leaves the state 
that can be represented as the squeezed two-photon states. 
The second one is the existence of the squeezed vacuum 
as the second term 
in Eq. (\ref{Ideal two photon subtracted state}). 
This corresponds to the term 
$\hat S \ket{0_+,0_-}$ $[=\hat S \ket{0,0}]$ 
in Eq. (\ref{Two photon subtracted state in biased modes}) 
[Eq. (\ref{Two photon subtracted state in unbiased modes})].  
This term allows one to tune the coefficients of superposition 
between the squeezed two-photon state and the squeezed vacuum. 
The second aspect makes a sharp contrast with 
a similar scheme with 
{\it nondegenerate two-mode squeezed vacuum} 
to generate two-photon state proposed in 
\,%
\cite{Nielsen_Molmer2007_PRA75_043801}, 
where such a term does not exist.

The state structure with the parameters 
$\Delta$ and the squeezing coefficient $\epsilon$ 
explained above allows one to control the bosonic interference 
in the time domain, 
and generate variety of quantum superposition states.

\section{Summary}
\label{Summary}

We studied non-Gaussian states generated 
by two-photon subtraction from a cw squeezed light source. 
A trigger photon click specifies  
a certain temporally localized mode 
in the remaining squeezed beam. 
The two-photon subtraction from 
overlappping squeezed packets generates 
temporally multiplexed 
Schr\"{o}dinger kittens with a designated time separation 
$\Delta$. 
They are generally an entangled states of even- and odd-number kittens, 
and the squeezed vacuum over two temporal modes. 
We have numerically studied these states 
by calculating the Wigner fuctions 
of the reduced states into each mode of the two kinds of sets. 
In the biased mode $\Psi_+(t)$ 
amplified kittens are produced 
in a region $\Delta\sim\zeta_0^{-1}$ 
where the time separation is comparable with 
the correlation time of squeezed packets.
This is due to constructive bosonic interference in the time domain.

As for furure perspective, 
it is interesting to consider ways of combining 
temporally multiplexed cat states 
with spatial degree of freedom 
to implement resource-effective quantum computation, 
and to generate useful nonclassical resources 
for metrology. 
It is also interesting to apply the similar 
operation with overlapping squeezed packets to 
other kinds of bosonic systems 
in other degrees of freedom 
such as the spatial and/or momentum domain.

\begin{acknowledgments}
The authors acknowledge M. S. Kim for helpful discussions. 
This work was supported by 
a MEXT Grant-in-Aid for Scientific Research (B) 19340115, 
and for Young Scientists (B) 19740253. 

\end{acknowledgments}


\appendix

\section{Mathematical formulae for two-phton subtraction}
\label{Appendix:two-phton subtraction}
%
%

In this Appendix we present methematical formulae 
to analyze the two-photon subtracted squeezed states 
in the cw scheme.

The trigger channel consists of on/off detectors 
with the the time resolution $T$ ($\sim1$\,ns). 
The trigger photon mode is assumed to be rectangular 
with the duration $T$ 
\beq\label{DFT functions}
\phi_k(t)
=
\left\{
\begin{array}{ll}
\displaystyle\frac{1}{\sqrt{T}} 
\mathrm{exp}(-i\displaystyle\frac{2\pi kt}{T}) 
& \quad 
 -\displaystyle\frac{T}{2}
  \le t \le 
  \displaystyle\frac{T}{2}, 
\\
0 
& \quad 
\mathrm{otherwise},
\end{array}\right. 
\eeq
Since $\zeta_0 T\ll1$ only the lowest mode $\phi_0(t)$ 
is excited in the photodetectors, 
and the weights of the other higher modes are negligible
\,\cite{Sasaki_Suzuki_PRA73_2006}.  
The trigger fields at $t_1$- and $t_2$-segments 
are defined by 
\beqa
\bar B
&\equiv&
\frac{1}{\sqrt T}\int_{t_1-T/2}^{t_1+T/2}dt \hat B(t)
\\
\bar C
&\equiv&
\frac{1}{\sqrt T}\int_{t_2-T/2}^{t_2+T/2}dt \hat C(t).  
\eeqa
%
%
%
%
%
%
Let $\bar{\Pi}_B$ and $\bar{\Pi}_C$ be 
the ``on" signal elements of 
the positive operator-valued measure (POVM) 
of the on/off detectors given in
\,\cite{Sasaki_Suzuki_PRA73_2006}.

The output state in path A conditioned by the ``on" signals 
is given by 
\beq\label{rho_A by two photon subtraction:Appendix}
\hat \rho_A
=
\frac{1}{P_{det}}
\tr_{BC} 
\left( \hat\rho_{ABC} \bar{\Pi}_B\otimes\bar{\Pi}_C \right) 
\eeq
where 
\beq\label{P_det:Appendix}
P_{det}
=
\tr_{ABC} \left( 
           \hat\rho_{ABC} \bar{\Pi}_B\otimes\bar{\Pi}_C
           \right) 
\eeq
is the detection probability. 
Here note that we discard the information of 
the ``off" signals.

This state is still a cw beam, consisting of multimodes. 
We are, however, interested in the two modes; 
the unbiased modes $\{ \Psi_{U1}(t), \Psi_{U2}(t)\}$ 
and the biased modes $\{ \Psi_+(t), \Psi_-(t)\}$.  
Denoting these sets as $\{\Psi_k(t)\}$  
we introduce the annihilation operators  
\beq\label{hat A_k:Appendix}
\hat A_k
\equiv 
\int_{-\infty}^{\infty} dt \hat A(t) \Psi_k(t).  
\eeq
Let the quadrature amplitude and phase operators be 
\beq\label{quadrature for beam A:Appendix}
\hat X_k
=
\frac{\hat A_k + \hat A_k^{\dagger}}{\sqrt2}, 
\quad 
\hat P_k
=
\frac{\hat A_k - \hat A_k^{\dagger}}{{\sqrt2}i},
\eeq
their vector notation be 
\beq\label{vector of quadratures for beam A:Appendix}
\hat {\mathbf{X}}_A
\equiv
\left(
  \begin{array}{c}
  \hat{X}_1 \\
  \hat{X}_2
  \end{array}
\right), 
\quad
\hat {\mathbf{P}}_A
\equiv
\left(
  \begin{array}{c}
  \hat{P}_1 \\
  \hat{P}_2
  \end{array}
\right), 
\eeq
and for beam B and C be 
\beqa\label{quadrature for beam B and C}
&\bar X_B
=\displaystyle
\frac{\bar B + \bar B^\dagger}{\sqrt2}, 
\quad 
&\bar P_B
=
\frac{\bar B - \bar B^\dagger}{{\sqrt2}i}
\\
&\bar X_C
=\displaystyle
\frac{\bar C + \bar C^\dagger}{\sqrt2}, 
\quad 
&\bar P_C
=
\frac{\bar C - \bar C^\dagger}{{\sqrt2}i}.
\eeqa

The three-beam squeezed state in the relevant modes 
is conveniently described by the characteristic function 
\beqa\label{Charac func of rho_ABC:def}
&&C(\hat\rho_{ABC};
    \mathbf{u}_A,\mathbf{v}_A,u_B,v_B,u_C,v_C)
\nonumber\\
&&=
\tr_{ABC} 
\Biggl\{
  \hat\rho_{ABC}
  \mathrm{exp}
  \biggl[i \bigl(  
           {}^t \mathbf{u}_A \hat {\mathbf{X}}_A 
         + {}^t \mathbf{v}_A \hat {\mathbf{P}}_A  
\nonumber\\
&&\qquad\qquad
         + u_B \bar X_B + v_B \bar P_B 
         + u_C \bar X_C + v_C \bar P_C 
           \bigr) \biggr]
\Biggr\}
\nonumber\\
&&=
\exp
\Bigl[
 -\frac{1}{4}{}^t 
\mathbf{u} \mathbf{\Gamma}(-\epsilon) \mathbf{u} 
\Bigr]
\exp
\Bigl[
 -\frac{1}{4}{}^t 
\mathbf{v} \mathbf{\Gamma}(\epsilon) \mathbf{v} 
\Bigr], 
\eeqa
where 
\beqa\label{Vector of u, v for ABC}
&\mathbf{u}
\equiv
\left(
  \begin{array}{c}
  \mathbf{u}_A \\
  u_{B} \\
  u_{C} 
  \end{array}
\right), 
\quad
&\mathbf{v}
\equiv
\left(
  \begin{array}{c}
  \mathbf{v}_A \\
  v_{B} \\
  v_{C}
  \end{array}
\right) 
\\
&\mathbf{u}_A
\equiv
\left(
  \begin{array}{c}
  u_1 \\
  u_2  
  \end{array}
\right), 
\quad
&\mathbf{v}_A
\equiv
\left(
  \begin{array}{c}
  v_1 \\
  v_2
  \end{array}
\right). 
\eeqa

\begin{widetext}
The covariance matrix is given by 
\beq\label{Matrix Gamma for ABC}
\mathbf{\Gamma}(\epsilon)
\equiv
\left(
  \begin{array}{ccc}
  \mathbf{\Gamma}_{AA}(\epsilon) 
    & \sqrt{1-R_1}\bm{\xi}_{B}(\epsilon)
    & \sqrt{R_1}\bm{\xi}_{C}(\epsilon) \\
  \sqrt{1-R_1}{}^t\bm{\xi}_{B}(\epsilon) 
    & 1+(1-R_1) b_{++}(\epsilon)
    & \sqrt{(1-R_1)R_1} b_{+-}(\epsilon) \\
  \sqrt{R_1}{}^t\bm{\xi}_{C}(\epsilon) 
    & \sqrt{(1-R_1)R_1} b_{+-}(\epsilon) 
    & 1+R_1 b_{++}(\epsilon) 
  \end{array}
\right) 
\eeq
\end{widetext}
where 
\beq\label{Matrix Gamma for A}
\mathbf{\Gamma}_{AA}(\epsilon) 
\equiv
R\mathbf{I}+(1-R)
\left(
  \begin{array}{ccc}
  M_{11}(\epsilon) & M_{12}(\epsilon) \\
  M_{21}(\epsilon) & M_{22}(\epsilon) 
  \end{array}
\right) 
\eeq
\beq\label{M_kl}
M_{kl}(\epsilon)
=\delta_{kl}
-
\frac{\epsilon \gamma_T}{\zeta(\epsilon)}
\int_{-\infty}^{\infty}dt\int_{-\infty}^{\infty}dt' 
\Psi_k(t)\Psi_l(t')e^{-\zeta(\epsilon)\vert t-t'\vert}  
\eeq
\beq\label{Vector of xi}
\bm{\xi}_{B}(\epsilon)
\equiv
\left(
  \begin{array}{c}
  \xi_1^B(\epsilon) \\
  \xi_2^B(\epsilon)  
  \end{array}
\right), 
\quad
\bm{\xi}_{C}(\epsilon)
\equiv
\left(
  \begin{array}{c}
  \xi_1^C(\epsilon) \\
  \xi_2^C(\epsilon)  
  \end{array}
\right) 
\eeq
\beq\label{xi_k^B}
\xi_k^B(\epsilon)
=
\sqrt{(1-R)R}
\frac{\epsilon \gamma_T}{\zeta(\epsilon)}
\sqrt{T}\int_{-\infty}^{\infty}dt 
\Psi_k(t) e^{-\zeta(\epsilon)\vert t-t_1\vert} 
\eeq
\beq\label{xi_k^C}
\xi_k^C(\epsilon)
=
\sqrt{(1-R)R}
\frac{\epsilon \gamma_T}{\zeta(\epsilon)}
\sqrt{T}\int_{-\infty}^{\infty}dt 
\Psi_k(t) e^{-\zeta(\epsilon)\vert t-t_2\vert} 
\eeq
\beqa\label{b_++ and b_+-}
b_{++}(\epsilon)
&=&-R\frac{\epsilon \gamma_T}{\zeta(\epsilon)}
\\
b_{+-}(\epsilon)
&=&b_{++}(\epsilon) e^{-\zeta(\epsilon)\Delta}.
\eeqa

The characteristic function of 
the two-photon subtracted squeezed state is given by 
\beqa\label{Characteristic func of rho_A 2:def}
&&C(\hat\rho_{A};\mathbf{u}_A,\mathbf{v}_A)
\nonumber\\
&&=
\frac{1}{P_{det}}
\tr_{ABC} 
\Bigl[
  \hat\rho_{ABC} \bar\Pi_B\otimes\bar\Pi_C
  e^{i ( {}^t \mathbf{u}_A \hat {\mathbf{X}}_A 
       + {}^t \mathbf{v}_A \hat {\mathbf{P}}_A )} 
\Bigr]
\nonumber\\
&&=
\frac{1}{(2\pi)^2 P_{det}}
\int du_B \int dv_B \int du_C \int dv_C
\nonumber\\
&&\qquad\times
C(\hat\rho_{ABC};\mathbf{u}_A,\mathbf{v}_A,u_B,v_B,u_C,v_C)
\nonumber\\
&&\qquad\times
C(\bar\Pi_B;-u_B,-v_B)
C(\bar\Pi_C;-u_C,-v_C). 
\eeqa
The characteristic function of the ``on" signal POVM element 
is given by 
\beqa\label{Characteristic func of bar Pi_B:def}
C(\bar\Pi_B;u_B,v_B)
&=&2\pi\delta(u_B)\delta(v_B)
\nonumber\\
&-&
\frac{e^{-\nu}}{\eta}
\mathrm{exp}
\Bigl[
  -\frac{2-\eta}{4\eta}
  \bigl(u_B^2+v_B^2\bigr) 
\Bigr] 
\eeqa
where $\eta$ is the overall detection efficiency, 
and $\nu$ is the fake trigger rate. 
Substituting it into 
Eq.~(\ref{Characteristic func of rho_A 2:def}) 
we have 
\beqa\label{Characteristic func of rho_A 2:expression}
C(\hat\rho_{A};\mathbf{u}_A ,\mathbf{v}_A ) 
&=&
\displaystyle\sum_{i=1}^4
\frac{{\cal N}^{(i)}}{P_{det}}
\exp
\Bigl[
      -\frac{ {}^t \mathbf{u}_A 
                   \mathbf{\Gamma}^{(i)}(-\epsilon) 
                   \mathbf{u}_A
             }{4}      
\Bigr]
\nonumber\\
&&\qquad\times
\exp
\Bigl[
      -\frac{ {}^t \mathbf{v}_A 
                   \mathbf{\Gamma}^{(i)}(\epsilon) 
                   \mathbf{v}_A
             }{4}      
\Bigr]
\eeqa
where 
\beqa
{\cal N}^{(1)}
&=&
1
\\
{\cal N}^{(2)}
&=&
-\frac{2e^{-\nu}}
{\eta\sqrt{b^{(2)}(-\epsilon)b^{(2)}(\epsilon)}}
\nonumber\\
{\cal N}^{(3)}
&=&
-\frac{2e^{-\nu}}
{\eta\sqrt{b^{(3)}(-\epsilon)b^{(3)}(\epsilon)}}
\nonumber\\
{\cal N}^{(4)}
&=&
\frac{4e^{-2\nu}}
{\eta^2\sqrt{b^{(4)}(-\epsilon)b^{(4)}(\epsilon)}}
\eeqa
\beqa
b^{(1)}(\epsilon)&=&1\\
b^{(2)}(\epsilon)&=&\frac{2}{\eta}R_1 b_{++}(\epsilon)\\
b^{(3)}(\epsilon)&=&\frac{2}{\eta}(1-R_1)b_{++}(\epsilon)\\
b^{(4)}(\epsilon)&=&b^{(3)}(\epsilon)b^{(2)}(\epsilon)
- (1-R_1)R_1 b_{+-}^2(\epsilon)
\eeqa
\beq
P_{det} = \sum_{i=1}^4 {\cal N}^{(i)}
\eeq
\beq
\Gamma_{kl}^{(i)}(\epsilon) 
=
R+(1-R)M_{kl}(\epsilon)
-\frac{j_{kl}^{(i)}(\epsilon)}{b^{(i)}(\epsilon)}
\eeq
\beqa
j_{kl}^{(1)}(\epsilon)&=&0\\
j_{kl}^{(2)}(\epsilon)&=&
      R_1\xi_k^C(\epsilon)\xi_l^C(\epsilon)\\
j_{kl}^{(3)}(\epsilon)&=&
      (1-R_1)\xi_k^B(\epsilon)\xi_l^B(\epsilon)\\
j_{kl}^{(4)}(\epsilon)&=&
      R_1 b^{(3)}(\epsilon)\xi_k^C(\epsilon)\xi_l^C(\epsilon)
\nonumber\\
      &+& (1-R_1) b^{(2)}(\epsilon)
           \xi_k^B(\epsilon)\xi_l^B(\epsilon)
\nonumber\\
      &-&  2(1-R_1)R_1 b_{+-}(\epsilon)
           \xi_k^B(\epsilon)\xi_l^C(\epsilon)
\eeqa

In real experiment 
the output state suffers from propagation loss, 
and 
imperfect efficiency of homodyne detection. 
Taking these factors into account by the effective 
transmission coefficient $\eta_H$ 
the observed state $\hat\rho_{A}'$ is characterized by 
\beqa\label{Characteristic func of measured rho_A 2:measured}
C(\hat\rho_{A}';\mathbf{u}_A ,\mathbf{v}_A ) 
&=&
\displaystyle\sum_{i=1}^4
\frac{{\cal N}^{(i)}}{P_{det}}
\exp
\Bigl[
      -\frac{ {}^t \mathbf{u}_A 
                   \mathbf{\Gamma}_H^{(i)}(-\epsilon) 
                   \mathbf{u}_A
             }{4}      
\Bigr]
\nonumber\\
&&\qquad\times
\exp
\Bigl[
      -\frac{ {}^t \mathbf{v}_A 
                   \mathbf{\Gamma}_H^{(i)}(\epsilon) 
                   \mathbf{v}_A
             }{4}      
\Bigr]
\eeqa
where 
\beq\label{eta_H}
\mathbf{\Gamma}_H^{(i)}(\epsilon) 
=(1-\eta_H)\mathbf{I}+\eta_H \mathbf{\Gamma}^{(i)}(\epsilon) 
\eeq
The output state also suffers from 
invasion of external noise photons from various parts, 
which degrades quantum correlations. 
This effect can be modeled by a mixing of 
thermal vacuum. 
Mathematically this can be taken into account 
by replacing the factor 
$\epsilon\gamma_T/\zeta(\epsilon)$ in 
Eqs. (\ref{M_kl}), (\ref{xi_k^B}),(\ref{xi_k^C}), 
and (\ref{b_++ and b_+-}) 
with $(\epsilon-\epsilon_x)\gamma_T/\zeta(\epsilon)$ 
where $\epsilon_x$ represents the degree of 
external noise photons.

The two-mode Wigner function is then given by 
\beqa\label{Two mode Wigner function}
W(\mathbf{x},\mathbf{p})
&=&
\frac{1}{(2\pi)^4}
  \int d\mathbf{u}_A d\mathbf{v}_A
  C(\hat\rho_{A}';\mathbf{u}_A ,\mathbf{v}_A ) 
  e^{i ( {}^t \mathbf{u}_A \mathbf{x}
       + {}^t \mathbf{v}_A \mathbf{p} )} 
\nonumber\\
&=&
\displaystyle\sum_{i=1}^4
   \frac{{\cal N}^{(i)}}{\pi^2 P_{det}}
   \cdot
   \frac{\exp\Bigl[-{}^t \mathbf{x} 
                    \mathbf{\Gamma}_H^{(i)}(-\epsilon)^{-1} 
                    \mathbf{x}
             \Bigr] }
        {\sqrt{|\mathbf{\Gamma}_H^{(i)}(-\epsilon)|} }
\nonumber\\
&&\qquad\qquad\quad\times
   \frac{\exp\Bigl[-{}^t \mathbf{p} 
                    \mathbf{\Gamma}_H^{(i)}(\epsilon)^{-1} 
                    \mathbf{p}
             \Bigr] }
        {\sqrt{|\mathbf{\Gamma}_H^{(i)}(\epsilon)|} }.
\eeqa

It is hard to depict the two-mode Wigner function 
numerically in a graph.  
One way to understand the state properties 
is to see the Wigner function reduced
to an appropriate single mode $\Psi_k(t)$ 
($k=U1$, $U2$, $+$, or $-$). 
It is given by 
\beq\label{Single mode Wigner function}
W(x,p)
=
\displaystyle\sum_{i=1}^4
   \frac{{\cal N}^{(i)}
         \exp\left[
               -\displaystyle
                \frac{x^2}{\Gamma_{Hkk}^{(i)}(-\epsilon)}
               -\displaystyle
                \frac{p^2}{\Gamma_{Hkk}^{(i)}(\epsilon) }
             \right] }
        {\pi P_{det} \sqrt{\Gamma_{Hkk}^{(i)}(-\epsilon)
                           \Gamma_{Hkk}^{(i)}(\epsilon)} }.
\eeq

The photon number in each mode can be calculated by 
the characteristic function by 
\beqa
\langle \hat n_k \rangle
&=&
\langle \hat A_k^\dagger \hat A_k \rangle
\nonumber\\
&=&
-\frac{1}{2}
 \Biggl[
    \biggl(
        \frac{\partial^2}{\partial u_k^2}
      + \frac{\partial^2}{\partial v_k^2}
    \biggr)
 C(\hat\rho_{A};\mathbf{u}_A ,\mathbf{v}_A ) 
 \Biggr]_{\mathbf{u}_A =\mathbf{v}_A=0}
\nonumber\\
&-&\frac{1}{2}
\nonumber\\
&=&
\frac{1}{4 P_{det}} \sum_{i=1}^4 {\cal N}^{(i)} 
\Biggl[
\Gamma_{kk}^{(i)}(-\epsilon)+\Gamma_{kk}^{(i)}(\epsilon)
\Biggr]
\nonumber\\
&-&\frac{1}{2}. 
\eeqa
This can be reduced to explicit expressions 
for small $\zeta_0 T\ll1$ as 
\beq
\langle \hat n_\pm \rangle
=
\frac{z^4 G_1^{(\pm)}(\Delta)+z^2 G_2^{(\pm)}(\Delta)}
{2(1\pm I_\Delta)
 [z^2 (1+ I_\Delta^2)+e^{-2\Delta}]} 
\eeq
where $z\equiv\zeta_0/\epsilon$ 
\beqa
G_1^{(\pm)}(\Delta)
&\equiv&
(1+ I_\Delta^2) g_\Delta^{(\pm)}
+2(1\pm I_\Delta) f_\Delta^{(\pm)2}
\\
G_2^{(\pm)}(\Delta)
&\equiv&
\psi(\Delta)^2 g_\Delta^{(\pm)} 
+2(1\pm I_\Delta)^3 
\nonumber\\
&& \pm
4(1\pm I_\Delta)\psi(\Delta) f_\Delta^{(\pm)2} 
\eeqa
with 
\beqa
f_\Delta^{(\pm)}
&\equiv&
\frac{3}{2}(1\pm I_\Delta)
\pm
\frac{\Delta^2}{2}e^{-\Delta}
\\
g_\Delta^{(\pm)}
&\equiv&
5(1\pm I_\Delta)
\pm
e^{-\Delta}(2\Delta^2+\frac{\Delta^3}{3}).
\eeqa
For the unbiased modes 
\beqa
\langle \hat n_{U1} \rangle
&=&\langle \hat n_{U2} \rangle
\nonumber\\
&=&
\frac{1}{2} 
[\langle \hat n_+ \rangle+\langle \hat n_- \rangle]. 
\eeqa

\end{document}